\documentclass[reprint, superscriptaddress, preprintnumbers, aps, prl]{revtex4-1}
\pdfoutput=1

\usepackage{setup}% all includes & macros
\DeclareRobustCommand{\NNLOJET}{\textsc{NNLOjet}\xspace}
\DeclareRobustCommand{\diff}[1]{\ensuremath{\mathrm{d}#1}}
\DeclareRobustCommand{\obs}{\ensuremathcal{O}}
\DeclareRobustCommand{\FS}{\ensuremath{F}}

\begin{document}

\preprint{CERN-TH-2018-272, IPPP/18/108, ZU-TH 45/18, LTH 1188}

\title{Jet production in charged-current deep-inelastic scattering to third order in QCD}%

\author{T.\ Gehrmann}%
 \affiliation{Department of Physics, University of Zürich, CH-8057 Zürich, Switzerland}%
\author{A.\ Huss}%
 \affiliation{Theoretical Physics Department, CERN, 1211 Geneva 23, Switzerland}%
\author{J.\ Niehues}%
 \affiliation{Institute for Particle Physics Phenomenology, Durham University,  Durham DH1 3LE, UK}%
\author{A.\ Vogt}%
\affiliation{Department of Mathematical Sciences, University of Liverpool, Liverpool L69 3BX, UK} 
\author{D.\ M.\ Walker}%
 \affiliation{Institute for Particle Physics Phenomenology, Durham University,  Durham DH1 3LE, UK}%

\date{\today}% It is always \today, today,
             %  but any date may be explicitly specified

\begin{abstract}
The production of jets in charged-current deep-inelastic scattering (CC DIS) 
probes simultaneously the strong and the electroweak sectors of the Standard Model; its
measurement  provides important information on the quark flavour structure of the proton. 
We compute third-order (\N3\LO) perturbative QCD corrections to this process, fully differential 
in the jet and lepton kinematics. We observe a substantial reduction in the theory uncertainty, to sub-percent level 
throughout the relevant kinematical range, thus enabling precision phenomenology with jet observables. 
\end{abstract}
\maketitle

Low-multiplicity electroweak processes at colliders are some of the most important benchmark processes for our understanding of Standard Model physics, allowing precision measurements of fundamental parameters and providing crucial tests of the theoretical framework and its practical application in precision calculations. 
In order for these theory predictions to be directly comparable to experimental data, 
they must be able to account for arbitrary (infrared safe) cuts on the final states produced, a requirement which also allows predictions of multiple-differential exclusive cross sections.  
This is in contrast to inclusive calculations, which yield results for the full phase space by using analytical techniques to integrate out final-state information; their 
comparison to experiment then requires ad-hoc 
extrapolations of data from the measured fiducial regions to the full phase space.  
 
Fully differential next-to-next-to-leading order (\NNLO) calculations are fast becoming the new theory benchmark for $2\rightarrow2$ scattering processes, with the completion of 
$\Pp\Pp\to\Pgg\Pgg$~\cite{twogamma}, 
$\Pp\Pp\to V\PH$~\cite{vh}, 
$\Pp\Pp\to V\Pgg$~\cite{vgamma,vgamma2}, 
$\Pp\Pp\to \Pqt\Paqt$~\cite{czakon}, 
$\Pp\Pp\to \PH+j$~\cite{hjet,ourhj},
% $\Pp\Pp\to \PW+j$~\cite{wjet,ourwj}, 
% $\Pp\Pp\to \PZ+j$~\cite{ourzj,zjet}, 
$\Pp\Pp\to V+j$~\cite{wjet,ourzj,zjet,ourwj}, 
$\Pp\Pp\to \Pgg+j$~\cite{mcfmgam},
% $\Pp\Pp\to \PZ\PZ$~\cite{zz}, 
% $\Pp\Pp\to \PW\PW$~\cite{ww}, 
% $\Pp\Pp\to \PZ\PW$~\cite{zw} and 
$\Pp\Pp\to VV$~\cite{zz,ww,zw} and 
$\Pp\Pp\to 2j$~\cite{2jnew} in proton--proton collisions, as well as the electron--positron process 
$\Pep\Pem \to 3j$~\cite{our3j,weinzierl3j}, the lepton--proton processes 
$\Pe\Pp\to \Pe+j$~\cite{abelof},
% $\Pe\Pp\to 2j$~\cite{disprl,ccdis} 
$\Pl\,\Pp\to \Pl'+2j$~\cite{disprl,ccdis}
and the related $2\to 3$ Higgs production processes in vector-boson fusion~\cite{vbfnnlo,vbfnnlo_ant,vbfnnloHH}. 
Here, $V$ denotes the massive gauge bosons $\PZ$ and $\PWpm$ and $j$ a reconstructed hadronic jet.
These calculations have been enabled by substantial developments~\cite{vbfnnlo,secdec,ourant,stripper,trocsanyi,qtsub,njettiness} of infrared subtraction methods for the handling of singular contributions that appear in all parton-level subprocesses. 

A limited number of processes have been evaluated inclusively beyond NNLO, notably neutral- and charged-current DIS~\cite{Photon_N3LO,CC_N3LO} as well as Higgs production in both gluon fusion~\cite{hn3lo,hy} and vector-boson fusion~\cite{vbfn3lo} to third order (\N3\LO), and 
$\Pep\Pem\to\:$hadrons to fourth order (\N4\LO)~\cite{epemn4lo}. 
These inclusive results can be used to perform differential calculations of closely related observables.
The Projection-to-Born (P2B) subtraction 
scheme~\cite{vbfnnlo,ourdis3} uses an inclusive calculation for a final state $\FS$ (fully differential in the 
variables of the Born-level kinematics) at a given perturbative order 
and an differential calculation of $\FS+\,$jet at one order lower in order to form a fully differential calculation. The scheme is unique in the fact that no new ingredients need to be calculated for the subtraction. 
It was first used in the case of Higgs production in vector-boson fusion ($\FS=\PH+2j$) at NNLO QCD~\cite{vbfnnlo}. 
The first application at \N3\LO was recently completed for the case of inclusive jet production in electromagnetic 
DIS ($\FS=\Pl+j$)~\cite{ourdis3}, where the inclusive structure functions calculated in~\cite{Photon_N3LO} were combined with the NNLO DIS di-jet calculation of~\cite{disprl} to form differential \N3\LO results. In an independent development~\cite{cieri}, the 
$q_\rT$-subtraction method~\cite{qtsub} was most recently extended to \N3\LO and applied to Higgs boson production 
at this order.

The general formula for a P2B cross section (multiply) differential in observable(s) \obs~is given by
\begin{equation}
  \frac{\diff{\sigma_\FS^{\N{k}\LO}}}{\diff{\obs}} = \frac{\diff\sigma_{\FS+j}^{\N{k-1}\LO}}{\diff{\obs}} - \frac{\diff\sigma_{\FS+j}^{\N{k-1}\LO}}{\diff{\obs_B}} + \frac{\diff\sigma_{\FS}^{\N{k}\LO,\mathrm{incl}}}{\diff{\obs_B}}\label{eqn:p2b_obs}\,,
\end{equation}
where a kinematic mapping uniquely assigns one Born-level observable \diff{\obs_B} for each \diff{\obs}:
\begin{equation}
  \diff{\obs} \xrightarrow{\text{P2B}} \diff{\obs_B}.\label{p2b_map}
\end{equation}
As a subtraction scheme, this is in effect using the matrix element itself as the counterterm to keep the integrand finite in all singular limits not already subtracted in the \N{k-1}\LO $\FS+j$ calculation. The integrated counterterm is exactly equivalent to the radiative contribution to the inclusive cross section which is mapped to the Born phase space during the analytic integration, such that when the three terms in (\ref{eqn:p2b_obs}) are combined, the fully differential cross section at \N{k}\LO is recovered. In this letter we use the NNLO calculation of di-jet production in charged current DIS implemented using antenna subtraction from~\cite{ccdis} alongside the inclusive \N3\LO charged current structure functions of~\cite{CC_N3LO} in order to produce for the first time differential distributions for single jet production in CC DIS ($\FS=\Pgn+j$). These corrections are all implemented in the parton-level Monte Carlo event generator \NNLOJET.

Charged-current deep-inelastic scattering (CC DIS) is a crucial process for our understanding of flavour content in parton distribution functions (PDFs), due to the preferential couplings of the \PW bosons to quarks dependent on their charge. CC DIS allows structure-function measurements in particular at high Bjorken-$x\gtrsim0.01$~\cite{h1cc,zeuscc}. With polarised incoming leptons it also allows precision tests of the chiral structure of the Standard Model, due to the linear dependence of the cross section on the polarisation fraction. If the proposed LHeC collider is constructed~\cite{LHeC}, precise predictions will become all the more relevant due to the vastly improved luminosity and kinematic reach compared to legacy HERA data.

The CC DIS process is equally relevant for current LHC predictions, in particular 
vector-boson fusion Higgs (VBF-Higgs) production.  In the structure-function approximation~\cite{han} 
the latter can be described well as ``double-DIS",  where each leg is constructed from independent DIS structure functions, with non-factorisable colour exchanges strongly suppressed by 
colour and kinematics. This relation is a strong motivation for improved NC and CC 
DIS predictions as many components are closely related. The \N3\LO inclusive cross section for single Higgs-boson~\cite{vbfn3lo} and double Higgs-boson~\cite{vbfn3loHH} production were 
calculated recently in this structure-function approximation. Combining these inclusive cross section with a calculation of VBF-Higgs production in association with a jet at NNLO using P2B will then allow the calculation of fully differential \N3\LO cross sections.

%% DIS kinematics
The kinematics of an inclusive charged-lepton CC DIS event takes the 
generic form
\begin{equation}\label{1}
  \Pp(P) + \Pl(k) \to \Pgn(k^{\prime}) + X(p_X),
\end{equation}
where $\Pp$ is the incoming proton, $\Pl$ the incoming charged lepton, $\Pgn$ the outgoing neutrino and $X$ a generic hadronic final state, with their corresponding momenta in brackets. The process is mediated by a $\PW$ boson of momentum 
$q=k-k^{\prime}$ with $Q^2=-q^2 > 0$, and can be fully described by the 
standard DIS variables
\begin{align}\label{2}
  s &= (P + k)^2\,, & 
  x &= \frac{Q^2}{ 2P\cdot q} \,, &
  y &= \frac{P \cdot q}{P \cdot k}\,.
\end{align}
Here $x$ is the usual Bjorken variable, and $y$ the scattering inelasticity 
(the fraction of the incoming lepton energy transferred to the proton in the proton rest frame).
Reconstructed jets can be further characterised through their transverse energy $E_{j}^T$, pseudorapidity $\eta_{j}$ and azimuthal angle $\varphi_{j}$ in the transverse plane.
The angle $\varphi_{j}$ is defined relative to the opposite direction of the lepton, i.e.\ $\varphi_{j}\equiv0$ for Born-level kinematics where the jet and lepton recoil back-to-back in the transverse plane.
As such, $\varphi_{j}$ only becomes non-trivial under additional radiation that escapes the jet clustering and constitutes a genuine DIS di-jet observable.

In the laboratory frame, the Born level kinematics of a single-jet CC DIS event can be fully reconstructed from the incoming beam energies and outgoing neutrino momentum, using momentum conservation:
\begin{align}
  p_{\text{in}, B}  &= xP\,, & 
  p_{\text{out}, B} &= xP + q \,.
\end{align}
Using this, one can define the unique map in (\ref{p2b_map}) from a final state of higher QCD multiplicity to the Born level, fulfilling the requirements for a consistent evaluation of \N3\LO jet production in DIS through P2B.

The ZEUS collaboration has measured inclusive jet distributions in the collision of $920~\GeV$ protons with polarised $27.6~\GeV$ electrons/positrons corresponding to a centre-of-mass energy of $\sqrt{s}=318.7~\GeV$~\cite{ZEUS}. The measurements were taken as functions of $x$, $Q^2$, leading-jet transverse energy $E_{j}^T$ and pseudorapidity $\eta_{j}$ for inclusive jet production. In the experimental analysis, the jets are $E_T$ ordered and clustered in the laboratory frame, applying the $k_T$-clustering algorithm in the longitudinally invariant mode. Data are presented for both $\Pep\Pp$ and $\Pem\Pp$ collisions, and are corrected for polarisation effects to give unpolarised cross sections.

% Double check these
In our calculation, the electroweak parameters are defined in the $G_{\mu}$-scheme, with $\PW$-boson mass $M_\PW=80.398~\GeV$, width $\Gamma_\PW=2.1054~\GeV$, $\PZ$-boson mass $M_\PZ=91.1876~\GeV$, coupling constant $\alpha=1/132.3384$ and Fermi constant $G_F=1.166\times 10^{-5}~\GeV^{-2}$. The number of massless flavours is five and contributions from massive top-quark loops are neglected. The calculations are performed using the NNPDF31 PDF set~\cite{nnpdf} with $\alphas(M_\PZ)=0.118$. We use the central renormalisation ($\muR$) and factorisation ($\muF$) scales $\muF^2=\muR^2=Q^2$. Scale variation uncertainties are estimated by varying $\muR$ and $\muF$ independently by factors of 0.5 and 2, but restricted to $0.5\leq\muR/\muF\leq2$.

Each event must pass the DIS cuts
\begin{align}\label{cuts}
  Q^2 &> 200~\GeV^2\,, &
  y   &< 0.9\,,
\end{align}
and the leading jet pseudorapidity must lie in the range $-1<\eta_{j}<2.5$ with minimum transverse energy $E_{j}^T > 14~\GeV$. The theory distributions are corrected for hadronisation and QED radiative effects using the multiplicative factors provided in~\cite{ZEUS}.
We validated the P2B implementation up to NNLO against the results using pure antenna subtraction in~\cite{ccdis} 
for $\Pep\Pp$ scattering for the above cuts, resulting in sub per-mille level agreement
(within numerical integration errors) for the NNLO contribution to the cross section: 
$-440.77 \pm 0.24~\fb$ (P2B)  versus $-440.82 \pm 0.30~\fb$ (antenna). 
When combined with the LO and NLO terms, we are confident that we have excellent agreement between the two methods for all choices of renormalisation and factorisation scales.

% Mention that PDFs and splitting functions for their evolution haven't caught up yet
%The PDF evolution is performed only to NNLO as PDFs with \N3\LO accuracy are not yet available.
\begin{figure*}[t]
  \centering
  \begin{tabular}{@{}cc@{}}  % \includegraphics[width=0.5\textwidth]{vardef}
  \includegraphics[scale=0.36]{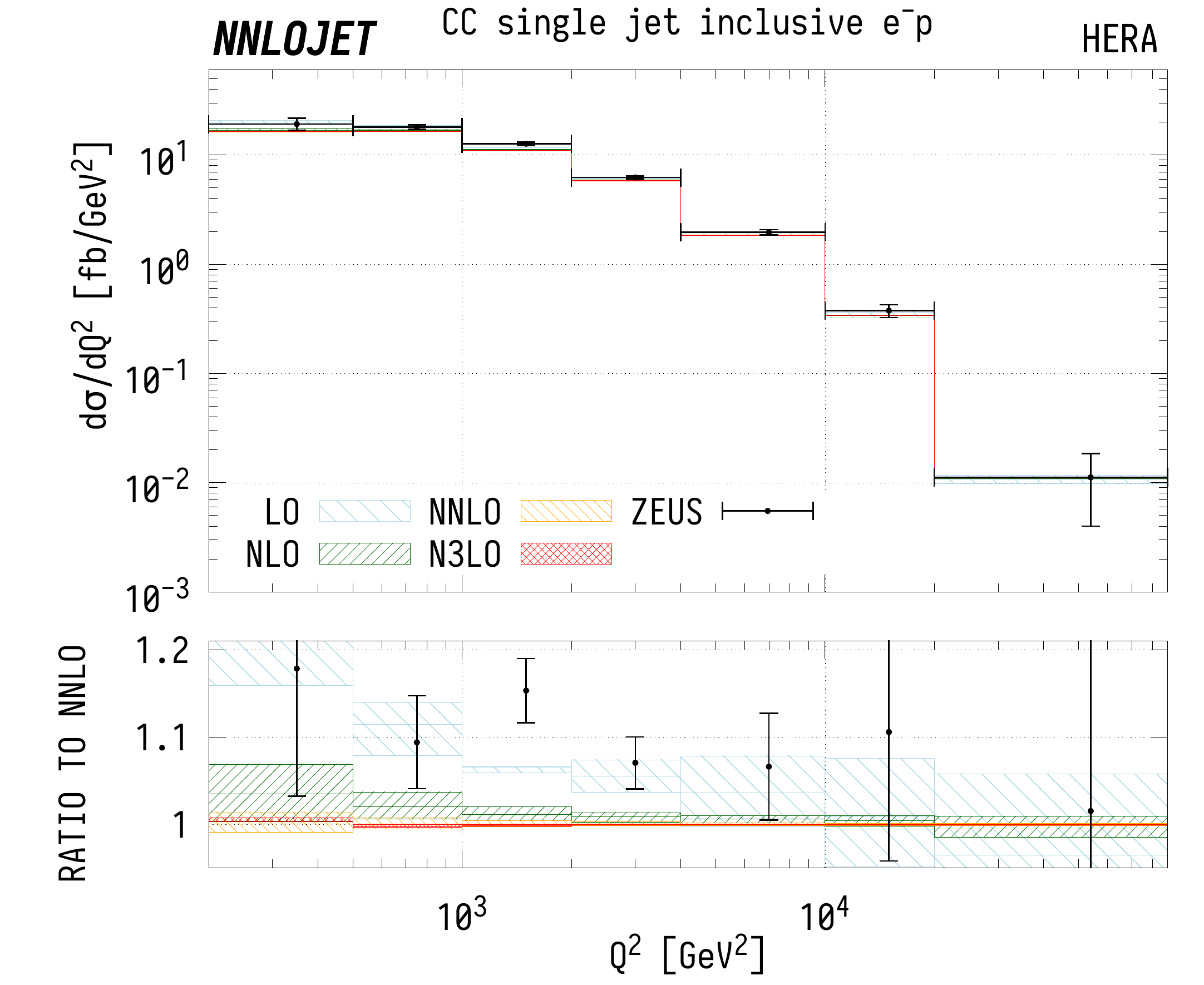} &
  \includegraphics[scale=0.36]{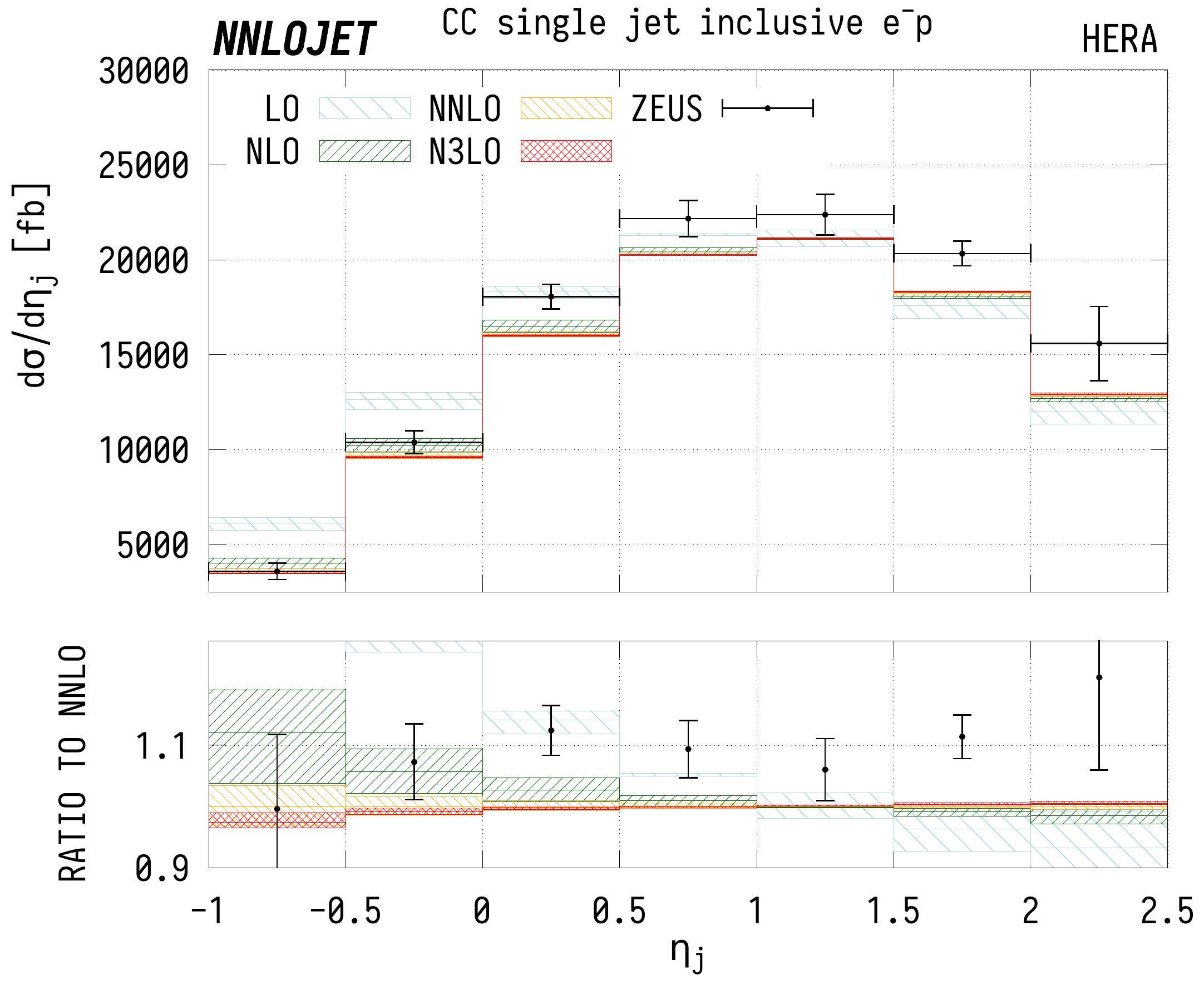} \\
  \includegraphics[scale=0.36]{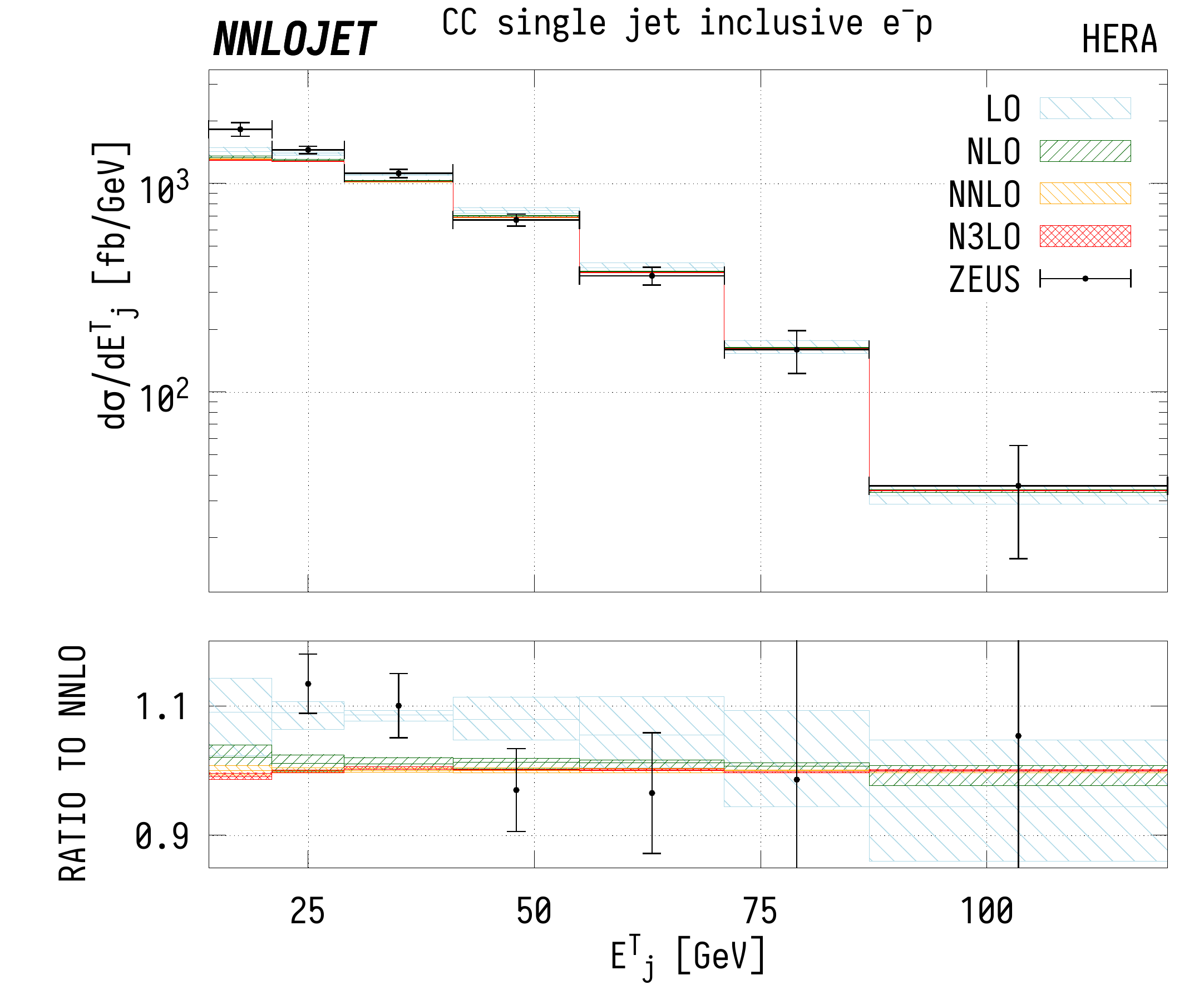} &
  \includegraphics[scale=0.36]{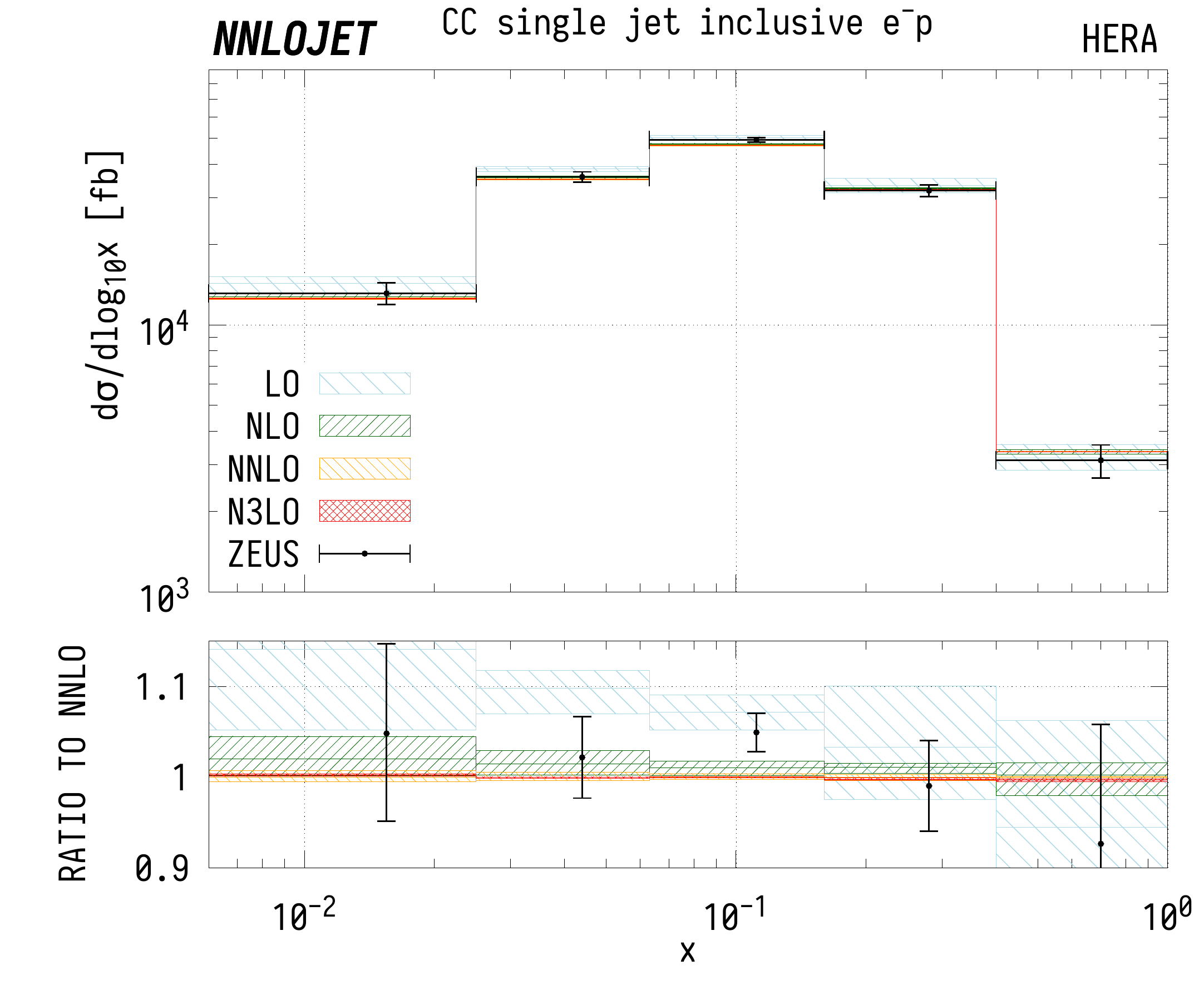} \\
  \end{tabular}
  \caption{Predictions at LO (blue left-hatched), NLO (green right-hatched), NNLO (orange left-hatched) and \N3\LO (red cross-hatched) are compared to ZEUS data from Ref. \cite{ZEUS} for $Q^2$, $\eta_{j}$, $E_{j}^T$ and Bjorken-$x$ for single jet production in $\Pem\Pp$ collisions. The bands correspond to scale uncertainties as described in the main text.}
  \label{fig:Pe-1}
\end{figure*}
It should be noted that the splitting functions for the PDFs are fully known 
only at \NNLO \cite{NNLOsplit}, for the status of the \N3\LO calculations see
\cite{N3LOsplit}, so for this calculation we have used NNLO PDFs. 
%For the $F_L$ and $F_3$ difference structure functions, we use the parameterised forms given in~\cite{CC_N3LO}. 
We do not expect that this will have any impact on the final results due to the small size of the overall correction.

\begin{figure*}[t]
  \centering
  \begin{tabular}{@{}cc@{}}  % \includegraphics[width=0.5\textwidth]{vardef}  
  \includegraphics[scale=0.36]{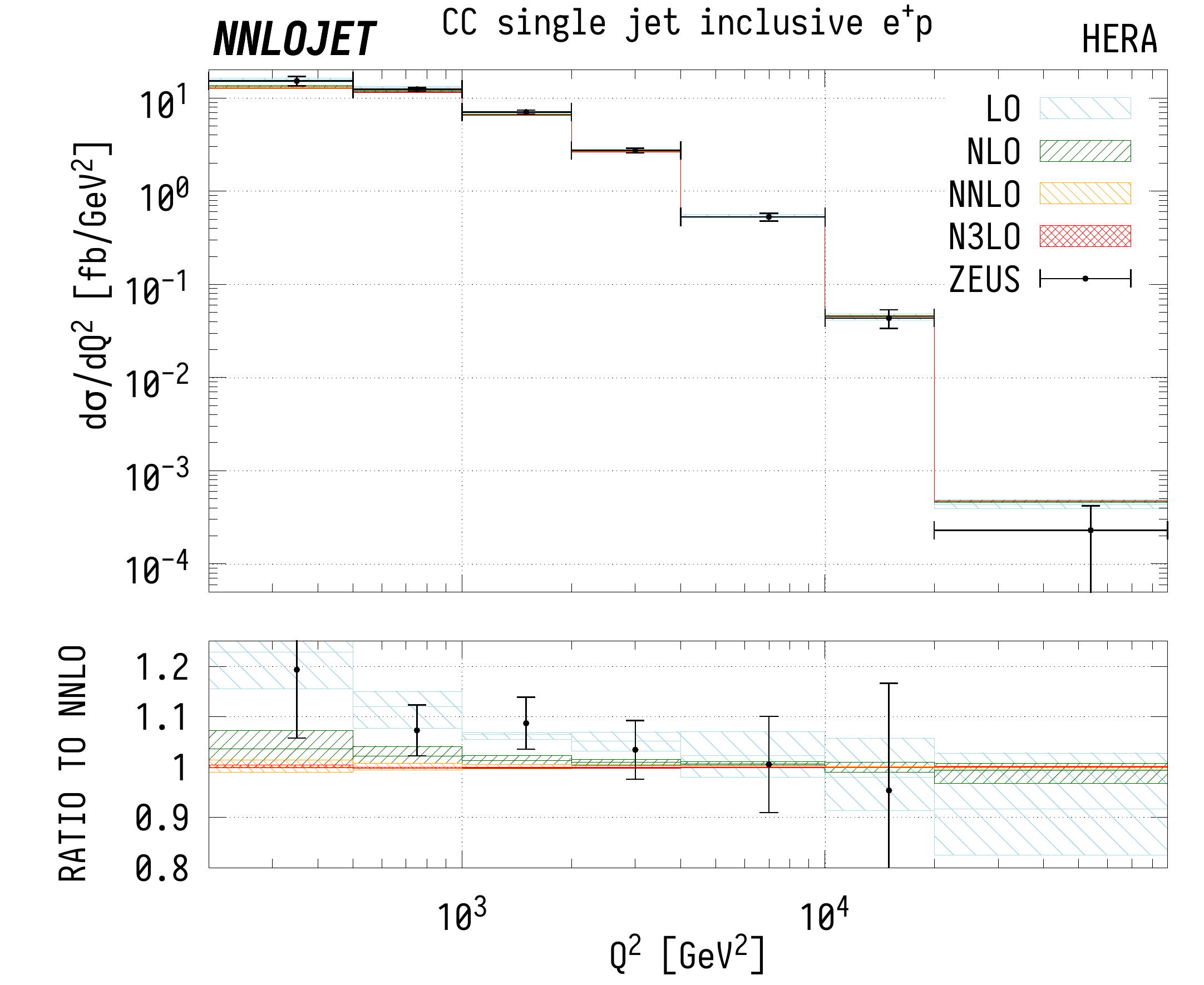} &
  \includegraphics[scale=0.36]{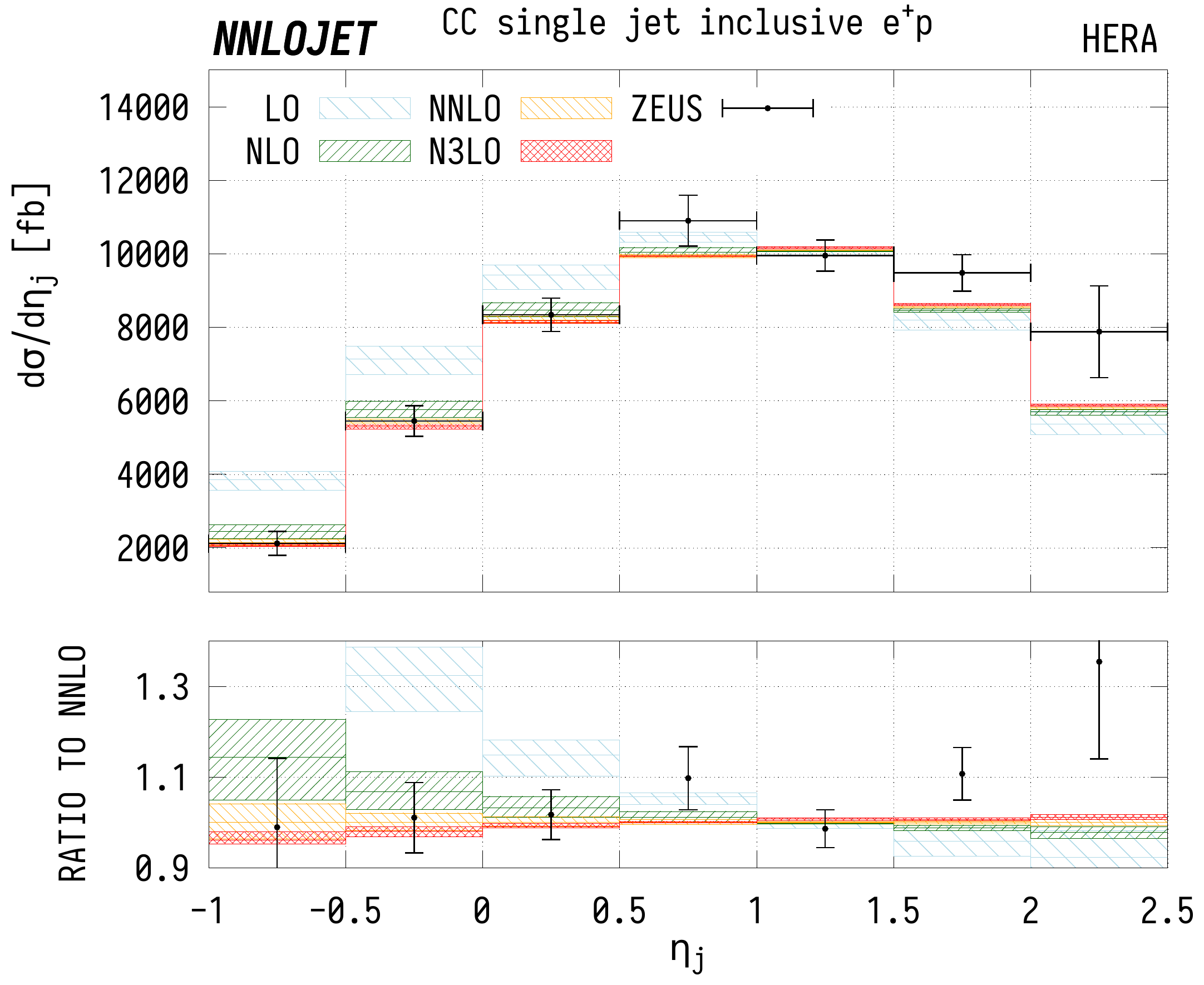} \\
  \includegraphics[scale=0.36]{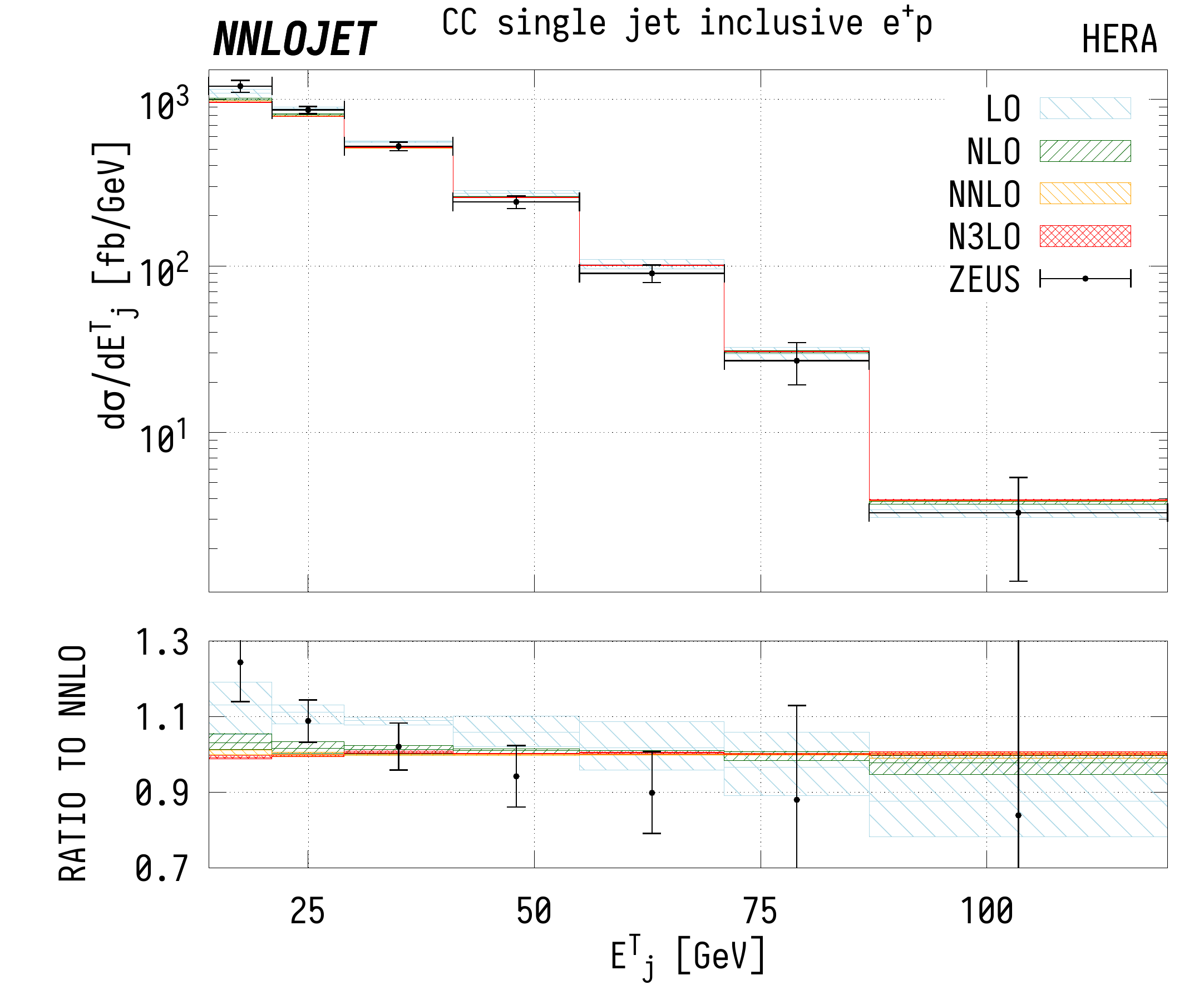} &
  \includegraphics[scale=0.36]{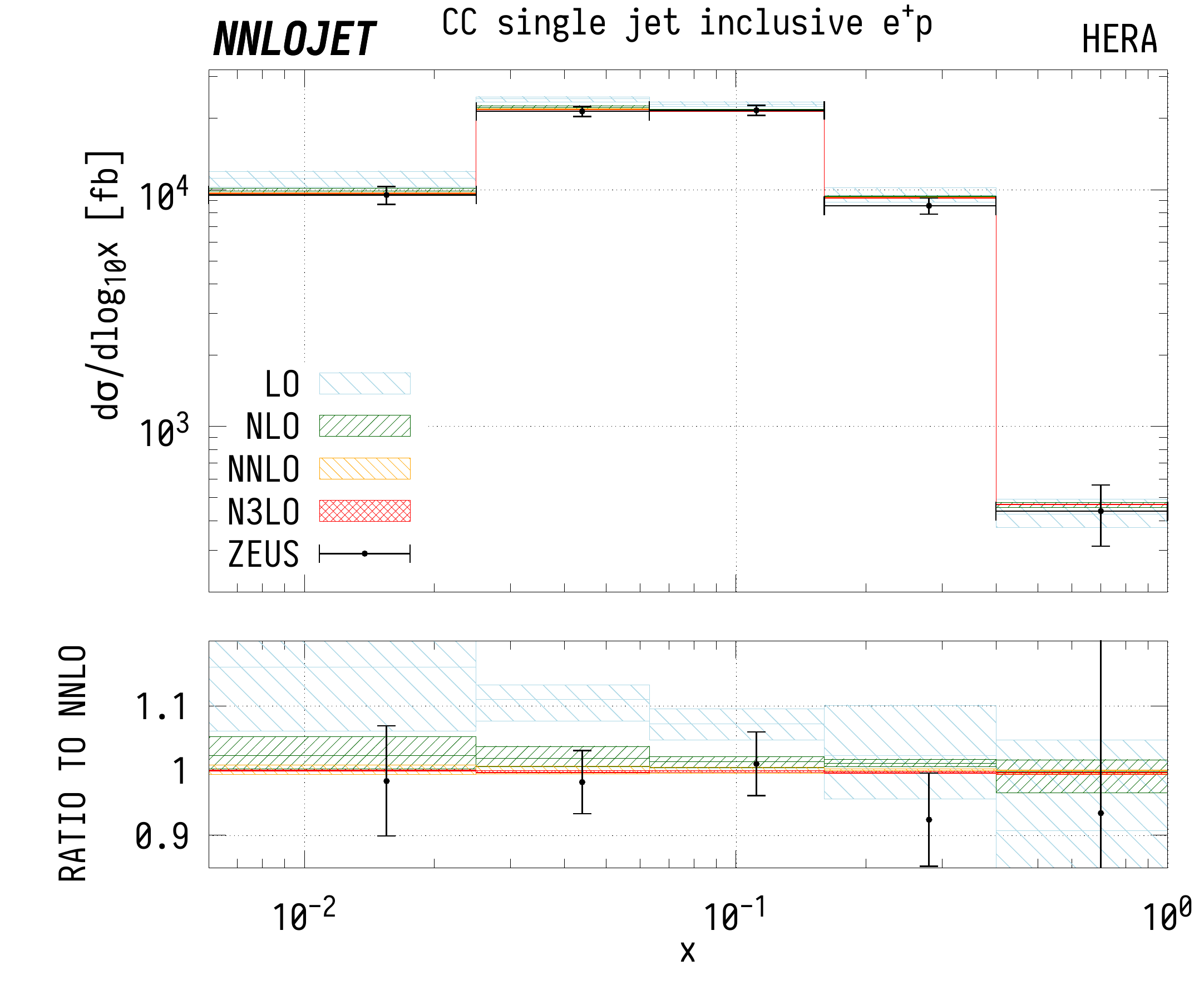} \\
  \end{tabular}
  \caption{Predictions at LO (blue left-hatched), NLO (green right-hatched), NNLO (orange left-hatched) and \N3\LO (red cross-hatched) are compared to ZEUS data from Ref. \cite{ZEUS} for $Q^2$, $\eta_{j}$, $E_{j}^T$ and Bjorken-$x$ for single jet production in $\Pep\Pp$ collisions. The bands correspond to scale uncertainties as described in the main text.}
  \label{fig:Pe+1}
\end{figure*}

%Comparisions to ZEUS single jet
A comparison of \NNLOJET predictions to ZEUS data for full cross sections differential in $Q^2$, $\eta_{j}$, $E_{j}^T$ and $x$ in single jet inclusive production in unpolarised $\Pem\Pp$ collisions is shown in Fig.~\ref{fig:Pe-1}. Corresponding results for unpolarised $\Pep\Pp$ collisions are shown in Fig.~\ref{fig:Pe+1}. In general, we find good agreement between theory and data, with overlapping scale uncertainty bands for NNLO  and \N3\LO predictions and a typical reduction in scale variation uncertainties going from NNLO to \N3\LO by a factor of two or better. Stabilisation of the perturbative QCD prediction can be observed for the first time below $\eta_j=0$  at this order. In the $Q^2$ distribution, the convergence of the prediction can now be seen in all bins, with the \N3\LO predictions contained fully within the NNLO scale variation bands. For low values of $x$ and $Q^2$, the predictions for $\Pem\Pp$ and $\Pep\Pp$ collisions begin to coincide as contributions from sea quarks and gluons inside the proton become dominant and differences between $\PWp$ and $\PWm$ exchange diminish. At larger values of $x$, valence-type 
quark distributions of the different charges determine the behaviour of the distributions.  
As noted already for the NNLO case in \cite{ccdis}, 
the agreement with data is systematically better for the $\Pep\Pp$ than for the $\Pem\Pp$ 
case, and can be traced back to a discrepancy in the $x$-distribution of $\Pem\Pp$ 
around $x\sim 0.15$, perhaps pointing to a PDF effect.  
 
%Conclusions
In this letter, we have presented the first fully differential calculation of single jet production in charged-current deep-inelastic scattering for both $\PWp$ and $\PWm$ exchanges at \N3\LO in QCD. 
We have applied our calculation to the kinematical situation relevant to the ZEUS experiment at HERA. The
\N3\LO predictions show perturbative stability throughout
the full kinematical range, and lead to a substantial reduction in scale uncertainty to sub per-cent level.   Together 
with the neutral-current results~\cite{ourdis3}, our calculation enables precision phenomenology 
with jet observables at a future LHeC collider~\cite{LHeC} and  constitutes an 
 important step to a fully differential \N3\LO calculation of vector-boson fusion Higgs production at
the LHC. 
%\bigskip

\section*{Acknowledgements}
The authors thank Xuan Chen, Juan Cruz-Martinez, James Currie, Rhorry Gauld, Aude Gehrmann-De Ridder, Nigel Glover, Marius H\"ofer, Imre Majer, Jonathan Mo, Tom Morgan, Joao Pires and James Whitehead for useful discussions and their many contributions to the \textsc{NNLOjet} code. 
This research was supported in part by the UK Science and Technology Facilities Council under contract ST/G000905/1, by the Swiss National Science Foundation (SNF) under contract 200020-175595, by the ERC Consolidator Grant HICCUP (No.\ 614577) and by the Research Executive Agency (REA) of the European Union under the  ERC Advanced Grant MC@NNLO (340983).

\end{document}